\documentclass[sigconf]{acmart}

\AtBeginDocument{%
  \providecommand\BibTeX{{%
    \normalfont B\kern-0.5em{\scshape i\kern-0.25em b}\kern-0.8em\TeX}}}

\copyrightyear{2021}
\acmYear{2021}
\setcopyright{rightsretained}
\acmConference[FAccT '21]{Conference on Fairness, Accountability, and Transparency}{March 3--10, 2021}{Virtual Event, Canada}
\acmBooktitle{Conference on Fairness, Accountability, and Transparency (FAccT '21), March 3--10, 2021, Virtual Event, Canada}
\acmDOI{10.1145/3442188.3445971}
\acmISBN{978-1-4503-8309-7/21/03}

\usepackage{color-edits}
\addauthor{sw}{blue}

\renewenvironment{quote}{%
   \list{}{%
     \leftmargin0.5cm   
     \rightmargin\leftmargin
   }
   \item\relax
}
{\endlist}

\usepackage{hyperref}
\hypersetup{
    colorlinks=true,
    linkcolor=blue,
    filecolor=magenta,      
    urlcolor=cyan,
}

\setcopyright{rightsretained} 
\begin{document}

\title{Value Cards: An Educational Toolkit for Teaching Social Impacts of Machine Learning through Deliberation}

\author{Hong Shen}
\email{hongs@cs.cmu.edu}
\affiliation{%
  \institution{Carnegie Mellon University}
  \streetaddress{5000 Forbes Ave}
  \city{Pittsburgh}
  \state{PA}
  \postcode{15213}
  \country{USA}
}

\author{Wesley H. Deng}
\email{wesley1016@berkeley.edu}
\affiliation{%
  \institution{University of California, Berkeley}
  \streetaddress{5000 Forbes Ave}
  \city{Berkeley}
  \state{CA}
  \postcode{94720}
  \country{USA}
}

\author{Aditi Chattopadhyay}
\email{aditic@andrew.cmu.edu}
\affiliation{%
  \institution{Carnegie Mellon University}
  \streetaddress{5000 Forbes Ave}
  \city{Pittsburgh}
  \state{PA}
  \postcode{15213}
  \country{USA}
}

\author{Zhiwei Steven Wu}
\email{zstevenwu@cmu.edu}
\affiliation{%
  \institution{Carnegie Mellon University}
  \streetaddress{5000 Forbes Ave}
  \city{Pittsburgh}
  \state{PA}
  \postcode{15213}
  \country{USA}
}

\author{Xu Wang}
\email{xwanghci@umich.edu}
\affiliation{%
  \institution{University of Michigan}
  \streetaddress{500 S State St}
  \city{Ann Arbor}
  \state{MI}
  \postcode{48109}
  \country{USA}
}

\author{Haiyi Zhu}
\email{haiyiz@cs.cmu.edu}
\affiliation{%
  \institution{Carnegie Mellon University}
  \streetaddress{5000 Forbes Ave}
  \city{Pittsburgh}
  \state{PA}
  \postcode{15213}
  \country{USA}
}

\renewcommand{\shortauthors}{Hong Shen et al.}

\newcommand\hs[1]{\textcolor{orange}{[Hong]: #1}}

\definecolor{lasallegreen}{rgb}{0.15, 0.66, 0.15}\newcommand\wdcomment[1]{\textcolor{lasallegreen}{[Wesley]: #1}}
\definecolor{chocolate(traditional)}{rgb}{0.48, 0.25, 0.0}\newcommand\xu[1]{\textcolor{chocolate(traditional)}{Xu: #1}}

\newcommand\hzcomment[1]{\textcolor{purple}{[Haiyi]: #1}}
\newcommand\accomment[1]{\textcolor{blue}{[Aditi]: #1}}

\begin{abstract}
Recently, there have been increasing calls for computer science curricula to complement existing technical training with topics related to Fairness, Accountability, Transparency and Ethics (FATE). In this paper, we present \textit{Value Cards}, an educational toolkit to inform students and practitioners the social impacts of different machine learning models via deliberation. This paper presents an early use of our approach in a college-level computer science course. Through an in-class activity, we report empirical data for the initial effectiveness of our approach. Our results suggest that the use of the Value Cards toolkit can improve students’ understanding of both the technical definitions and trade-offs of performance metrics and apply them in real-world contexts, help them recognize the significance of considering diverse social values in the development and deployment of algorithmic systems, and enable them to communicate, negotiate and synthesize the perspectives of diverse stakeholders. Our study also demonstrates a number of caveats we need to consider when using the different variants of the Value Cards toolkit. Finally, we discuss the challenges as well as future applications of our approach.

\end{abstract}





\maketitle

\section{Introduction}
Machine learning-based  decision-making systems have been deployed to address many high-stakes problems in our society, such as college admissions \cite{o2016weapons}, loan decisions \cite{Yerak2019}, and child maltreatment prediction \cite{brown2019toward}, raising many social and ethical concerns. Even with the best of intentions, ML development teams – constrained by limited team diversity and lack of necessary training – often struggle to fully comprehend the complicated and nuanced social issues embedded in many contemporary AI systems \cite{west2019discriminating,costanza2020design, holstein2019improving}. 

The need to teach computer science students -- and by extension, future practitioners -- the social impacts of the machine learning systems they are going to build has become urgent. As a result, there has been an increase in calls for computer science departments to complement existing technical training with issues related to Fairness, Accountability, Transparency and Ethics (FATE) \cite{leidig2020acm,singer2018tech,o2017ivory}. Indeed, we have witnessed a growing number of computer science departments seeking to infuse FATE topics into their curricula \cite{fiesler2020we}.

To date, a lot of work that focused on Fairness and Explainable AI has sought to develop \textit{technical} solutions in the form of toolkits and systems to help ML practitioners better comprehend, evaluate and debias their machine learning models (e.g., \cite{bellamy2019ai, wexler2019if}). However, the growing impact of algorithmic systems in our society has necessitated the need for more research efforts to be devoted to cultivating deeper understanding of the diverse and potentially competing \textit{social} values embedded in these systems (e.g., \cite{zhu2018value, lee2019webuildai,Shen2020CM}). 

\begin{figure}[h]
  \centering
  \includegraphics[width=\linewidth]{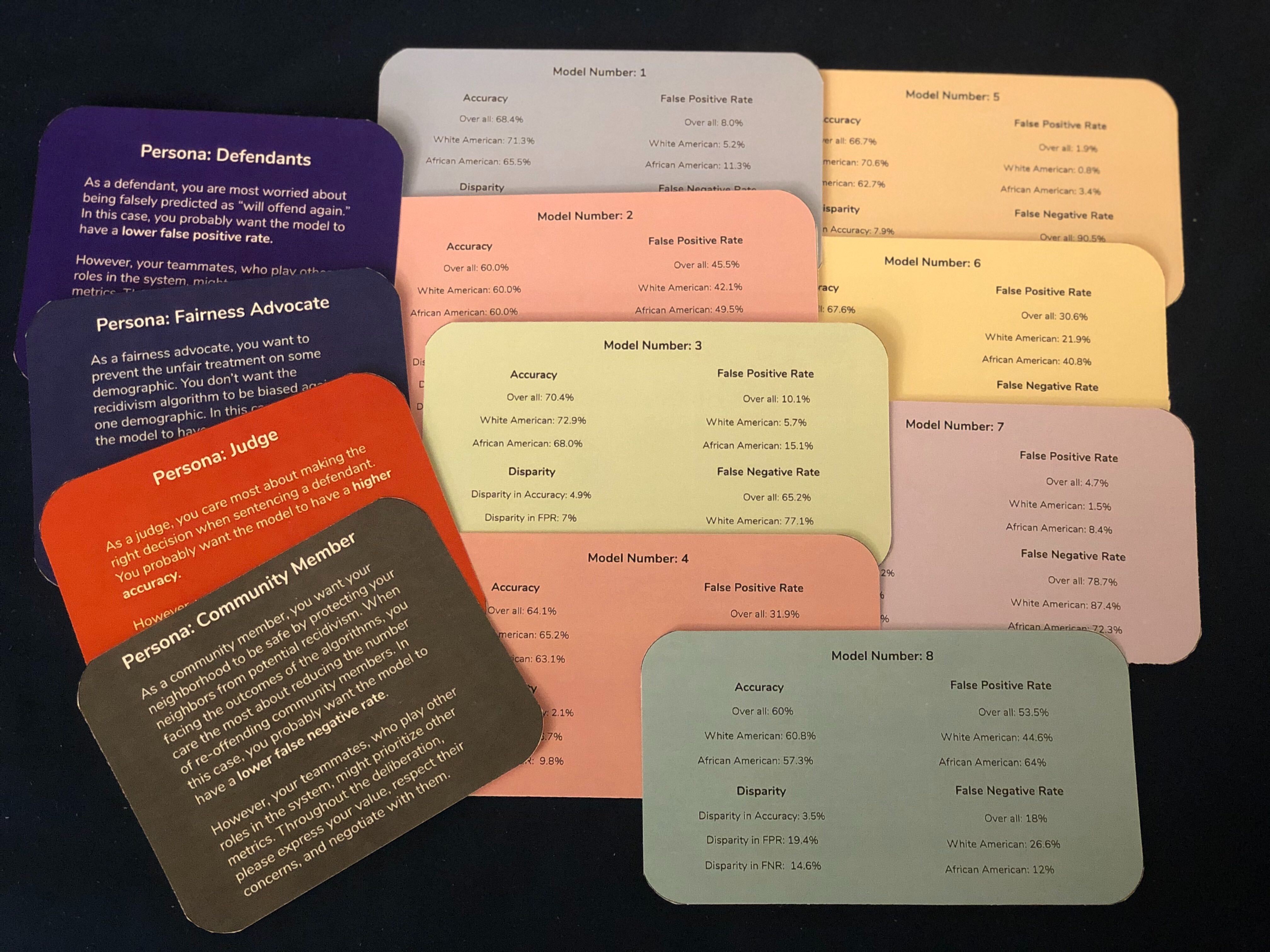}
  \caption{A snapshot of the Value Cards Toolkit, including eight Model Cards and four Persona Cards.}  
    \vspace{-10pt}
  \label{tab:value card}
\end{figure}

Our work contributes to this line of research. In this work, we present Value Cards (Figure~\ref{tab:value card}), a deliberation-driven toolkit for bringing computer science students and practitioners the awareness of the social impacts of machine learning-based decision making systems.
\footnote{We include all eight Model Cards, four Persona Cards, and three Checklist Cards in Appendix. The Value Cards toolkit, including all the related teaching materials, is also available at https://wesdeng.github.io/ValueCards/} Specifically, Value Cards encourages the investigations and debates towards different ML performance metrics and their potential trade-offs. Instead of viewing human values in AI systems as \textit{individual dilemmas} that can be calculated as aggregations of \textit{individual preferences}, we foreground the importance of negotiating \textit{social} values and making \textit{collective} decisions via deliberation \cite{zhu2018value,Shen2020Publics,Shen2020CM}. The Value Cards toolkit uses three different artifacts -- the Model Cards, the Persona Cards, and the Checklist Cards -- to better facilitate and scaffold the deliberation process. In particular, the Model Cards describe a set of machine learning models, which capture the inherent trade-offs in the design of a machine learning-based decision making system. The Persona Cards characterize different potential stakeholders and their prioritized values and interests in the system, with each corresponding to a number of specific Model Cards. The Checklist Cards enumerate a number of social and technical considerations to better facilitate and scaffold the deliberation process. Instructors can adjust and tailor the components as they need for their specific domains or settings. 

In this paper, we document an early use of the Value Cards toolkit in a college level computer science course. In our class activity, we test four variants of the toolkit: Model Cards only, Model Cards + Checklist Cards, Model Cards + Persona Cards, Model Cards + Checklist Cards + Persona Cards. Our evaluation shows that the use of Value Cards in general improves students’ understanding of the technical definitions and the trade-offs between different performance metrics, and their implications in the context chosen for the study: recidivism prediction. It helps students recognize the importance of considering diverse social values in the development and deployment of machine learning systems. It also enables them to communicate, negotiate, and synthesize the perspectives of diverse stakeholders. Our results also reveal a number of pros and cons that need to be considered when using the different variants of the Value Cards approach.  Our contributions are three-fold:
\begin{itemize}
\item First, we introduce a deliberation-driven toolkit -- the Value Cards -- to help computer science students and potential future practitioners better understand the social impacts of machine learning models. 

\item Second, we document an early use of our approach in a college-level computer science course and conduct an empirical investigation of the initial effectiveness of our approach in an educational setting. 

\item Third, we discuss the challenges of using different variants of the toolkit as well as future applications of our approach. 

\end{itemize}

\section{Related Work}
In this section, we outline relevant work in two areas. First, we survey the existing techniques, toolkits, and systems related to issues of FATE in the fair ML literature, and describe how our work is positioned in this space. Next, we present an overview of existing work on educational efforts around teaching FATE in computer science, a field that has received increasing attention in FAccT and describe how our work contributes to this emerging line of research.

\subsection{Developing techniques, toolkits and systems for FATE in Machine Learning}
Recently, there have been increasing concerns about algorithmic bias and unfairness in AI systems (e.g., see \cite{barocas2016big, eubanks2018automating, angwin2016machine}). As a result, significant effort in the fair ML community has been devoted to the development of algorithms, toolkits, and systems to aid ML development teams in assessing and addressing potential biases (e.g., see \cite{chen2018my,holstein2019improving,angell2018themis,galhotra2017fairness}). 

For example, Google's People + AI Research group (PAIR) developed the open-source ``What-if'' tool \cite{wexler2019if} to help practitioners who are not formally trained in machine learning visualize the effects of fairness metrics. Microsoft has also developed the Fairlearn toolkit (fairlearn.github.io) based on the work of \cite{AgarwalBD0W18, AgarwalDW19} and IBM has developed the AI Fairness 360 toolkit \cite{bellamy2019ai} to help ML developers assess and mitigate certain kinds of harms in ML models. More recently, researchers at Google developed ML-fairness-gym \cite{10.1145/3351095.3372878} to simulate the potential long-term impacts of deploying machine learning-based decision systems in social environments, offering ML practitioners the chance to conduct a dynamic analysis instead of a single-step or static one.

A different line of research has explored the use of detailed and multidisciplinary documentation techniques to enhance transparency in model and data reporting. For example, the Model Card approach \cite{mitchell2019model} details information such as the model type, intended use cases, performance characteristics of a trained ML model. Datasheets \cite{gebru2018datasheets} focuses more on clarifying the characteristics of the data feeding into the ML model. Drawing from approaches in document collection practices in archives, Jo and Gebru \cite{jo2020lessons} discuss five data collection and annotation approaches that can inform data collection in sociocultural machine learning systems. 

Our work takes a complementary angle, looking at helping computer science students and practitioners better understand the social impacts of machine learning. Inspired by Value Sensitive Algorithm Design \cite{zhu2018value}, a design process seeking to better incorporate stakeholder values in the creation of algorithmic systems, we introduce the Value Cards -- a deliberation-driven toolkit to help computer science students and practitioners better comprehend, negotiate, and reflect on those diverse and oftentimes competing social values in machine learning-powered algorithmic decision making systems. 

\subsection{Teaching FATE in Computer Science}
While teaching ethics and social responsibility has a long history in computer science education (e.g., \cite{martin1996implementing,nielsen1972social}), the recent widespread deployment of machine learning-powered algorithmic decision making systems in many high-stakes social domains has lead to a growing attention to issues related with Fairness, Accountability, Transparency and Ethics (FATE) \cite{leidig2020acm, singer2018tech,o2017ivory}. 

Indeed, we have witnessed an increasing number of computer science departments seeking to infuse topics related with FATE into their curricula. Recently, Fiesler et al. \cite{fiesler2020we} presented a qualitative analysis of 115 syllabi from university technology ethics courses and noted that there is a great variability in terms of instructors, topics and learning outcomes across different institutions. Past research in this space has offered important case studies and techniques, for example, using immersive theater \cite{skirpan2018quantified}, games  \cite{brinkman2017code}, and science fiction \cite{burton2018teach} for ethics pedagogy in computer science. More recently, Reich et al. \cite{10.1145/3328778.3366951} documented a curricular experiment at Stanford University, which combined philosophy, political science, and CS in teaching computer ethics. They found that compared with separating ethics and tech training, students resonated strongly with this multidisciplinary approach. Based on their experience in teaching FATE/Critical Data Studies (CDS) topics at University of Sheffield (UK) Information School, Bates et al. \cite{bates2020integrating} discussed a series of challenges educators face for deeper integration FATE topics into the existing curriculum. 

Similarly, Saltz et al. \cite{saltz2019integrating} specifically reviewed the current state of ML ethics education via an analysis of course syllabi. Their analysis demonstrated the need for ethics topics to be integrated within existing ML coursework, rather than stand-alone ethics courses. They also discussed a few novel examples of practically teaching ML ethics without sacrificing core course content. 

Our work contributes to this fast growing line of research. In this paper, we developed and evaluated a novel deliberation-driven toolkit -- the Value Cards -- to facilitate the education of FATE issues within existing CS coursework, rather than as separated ethics classes. Our approach strives to complement existing in-class technical training with informed understanding of the social impacts of machine learning-based decision making systems. It has the potential to be adopted in a wide variety of settings. 

\section{Design of The Value Cards Toolkit}
\subsection{General Design Rationale}
In this section, we describe the general design, rationale and objectives of the Value Cards toolkit. Inspired by the Envision Cards \cite{friedman2012envisioning} in  Value Sensitive Design \cite{friedman1996value,friedman2008value}, we consider the Value Cards as a versatile toolkit that can be used to facilitate the deliberation of different -- and often competing -- social values embedded in a variety of machine learning-based algorithmic systems. 

The core of the Value Cards approach is \textbf{a deliberation process}. Instead of viewing human values in AI systems as \textit{individual dilemmas} that can be calculated as aggregations of \textit{individual preferences}, we foreground the importance of \textit{social} values and \textit{collective} decision making via deliberation \cite{Shen2020Publics, zhu2018value}. Deliberation refers to an approach to politics in which members from the general public are involved in collective decision-making through the exchange of ideas and perspectives via rational discourse \cite{cavalier2011approaching}. Moreover, deliberation and discussion-based approaches have demonstrated benefits for different aspects of student learning, including conceptual learning, especially learning of difficult content \cite{azmitia1993friendship, schellens2007scripting}, acquisition of argumentative knowledge \cite{weinberger2010learning, wang2017contrasting}, and perspective taking \cite{wang2020practice}. We anticipate that through deliberation, participants may have the opportunity to understand each other's perspectives, challenge one another to think in new ways, and learn from those who are most adversely affected. 

Prior work in computer-supported cooperative work (CSCW) and Learning Sciences has shown the success of various strategies to support effective deliberation and discussion. Scaffolding discussion with scripts \cite{o1992scripted, dillenbourg2002over, weinberger2005epistemic} has demonstrated benefits on student learning in a variety of contexts \cite{azmitia1993friendship, schellens2007scripting, weinberger2010learning, wang2017contrasting}. For collaborative problem-solving and decision-making, a \textit{script} is a set of instructions regarding to how the group members should interact and how they should solve the problem together. One categorization of collaborative learning scripts divide them into social scripts, which structure how learners interact with each other, and epistemic scripts, which specify how learners work on a given task \cite{weinberger2005epistemic}. 

In the design of Value Cards toolkit, we instantiate the idea of the \textit{script} with three different artifacts 
-- the Model Cards, the Persona Cards, and the Checklist Cards -- to scaffold and facilitate the deliberation process. 
The \textbf{Model Cards} describe a set of machine learning models that capture the inherent trade-offs in the development of machine learning application. The \textbf{Persona Cards} depict different potential stakeholders and their prioritized values and interests. The \textbf{Checklist Cards} enumerate a number of social and technical considerations to better scaffold the deliberation process. Instructors can adjust the components as they need for their specific problem domains or settings. In our design, the Persona Cards are one form of social script, and both the Model Cards and the Checklist Cards are forms of epistemic script.

\subsection{Learning Objectives}
Taken as a group, the Value Cards toolkit is designed specifically to achieve the following three \textbf{learning objectives}. 
\begin{itemize}
    \item Understand the technical definitions and trade-offs of performance metrics in machine learning, and apply them in real-world contexts.
    \item Understand the importance of considering diverse stakeholders' perspectives in the development and deployment of machine learning systems.
    \item Being able to communicate, negotiate, and synthesize the perspectives of diverse stakeholders when algorithmic decisions have consequential impacts.
\end{itemize}

First, as past literature \cite{saltz2019integrating} has pointed out, current ethics education in ML tend to separate the technical training from the ethical training, here we offer an integrated toolkit to help students understand the technical definitions and trade-offs of performance metrics, and apply them in real-world contexts. In doing so, our approach has the potential to bridge the gap between model development and real-world application \cite{veale2018fairness}.

Second, as Yang et al. \cite{yang2018grounding} point out that there is a tendency among practitioners who are not formally trained in ML to overly prioritize accuracy over other system criteria in model building -- ``many perceived percentage accuracy as a sole measure of performance, thus problematic models proceeded to deployment.'' The Value Cards toolkit is designed to tackle this challenge by educating students on  the significance of considering diverse stakeholders’ perspectives in the development and deployment of ML systems.

Third, past literature in HCI also suggests that there is a lack of communication skills among developers when facing design-related tasks, which might result in poorly designed systems that can further disadvantage already-marginalized populations \cite{oleson2020computing}. The Value Cards toolkit also strives to enable computer science students to better communicate, negotiate, and synthesize the perspectives of diverse stakeholders when algorithmic decisions have consequential impacts. 

We next offer a ``proof of concept'' example by illustrating each component in our toolkit in the context of recidivism prediction.

\subsection{Domain and Dataset: Recidivism Prediction}
We used recidivism prediction as the context for exploring the general research problem of helping computer science students understand the societal aspect of machine learning-based algorithmic systems. Recidivism prediction is a high-stakes social context where algorithms have been increasingly deployed to help judges assess the risk of recidivism. As a contentious social domain that involves diverse and competing social and political interests, this topic has also generated one of the most controversial cases so far in the debates around fairness and machine learning \cite{angwin2016machine}. 

Following \cite{yu2020keeping}, we recreated a recidivism prediction tool using a data set provided by ProPublica \cite{larson2016we}. For the purpose of this study, we focused on two demographic groups, i.e., African American and White populations and created a balanced data set, which resulted in a data set of 3,000 defendants (1,500 White
defendants and 1,500 African American defendants). 

\subsection{The Model Cards}
\begin{figure}[h]
  \centering
  \includegraphics[width=\linewidth]{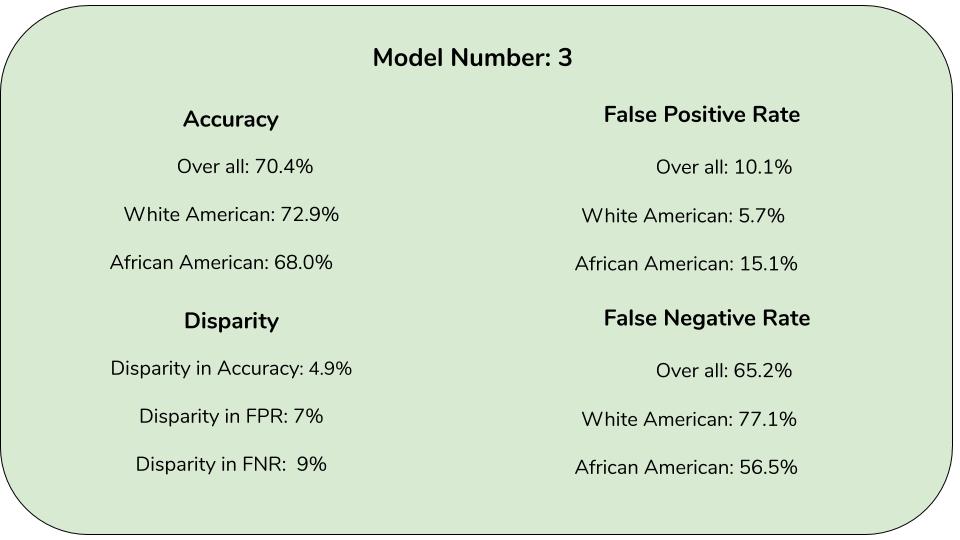}
  \caption{A Model Card.  }
  \Description{todo}
  \label{tab:model}
\end{figure}

Inspired by Mitchell et al. \cite{mitchell2019model}, the first set of artifacts in our toolkit is the Model Cards. Each Model Card describes one machine learning model by showing its performance metrics in aggregate and across different demographic groups (see Figure~\ref{tab:model} for an example). Collectively, the Model Cards capture the trade-offs across different metrics in a machine learning-based decision making system. 

The aim of the Model Cards is to both describe the performance metrics and to capture the inherent and often implicit trade-offs across different metrics in a machine learning-based decision making system. Optimizing for multiple system criteria is a tricky task: optimizing one criterion often leads to poor performance on others. The use of the Model Cards is to specifically capture, scaffold and communicate those trade-offs to the readers.  

In the case of recidivism prediction algorithm, following Yu et al. \cite{yu2020keeping}, we selected four sets of performance metrics, namely, accuracy (fraction of correct predictions), false positive rate (fraction of people who are falsely predicted to re-offend, among those who do not re-offend), false negative rate (fraction of people who are false predicted to not re-offend, among those who re-offend), and the disparities of these three measures between the two demographic groups: African American and White defendants.

By adopting the Lagrangian-based methods from \cite{AgarwalBD0W18, KearnsNRW18}, 
we generated a family of predictive models that exhibit a wide range of trade-offs between the four different system criteria outlined above. Next, given a family of models, we selected a subset of eight models approximately mapping back to the stakeholder valaues identified in the Persona Cards, as discussed below. We created eight Model Cards for these eight models. 

\begin{figure}[h]
  \centering
  \includegraphics[width=\linewidth]{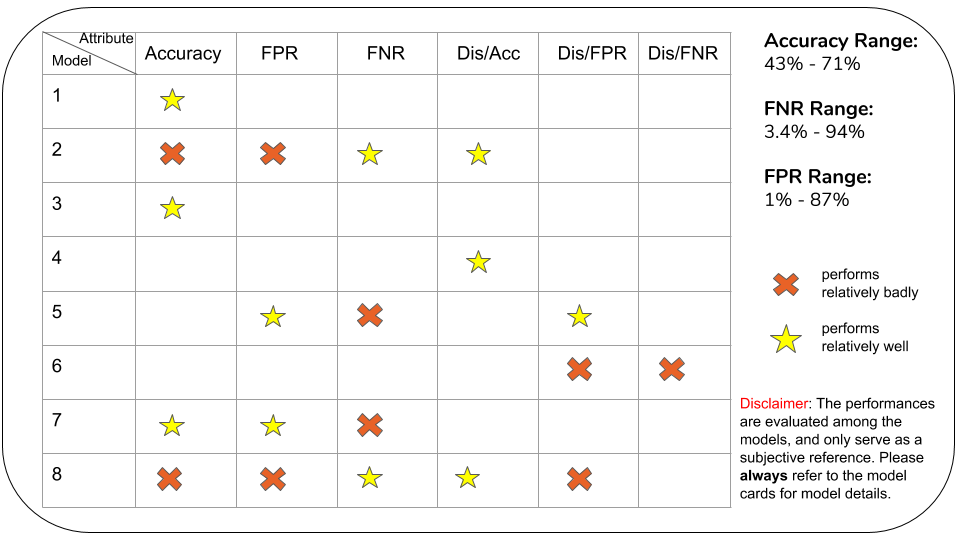}
  \caption{An overview table of the performance metrics across all the eight model cards we offered in the case study. We use golden star and red cross to indicate the relative metric performances. We also provide the percentage range of the metric performance and a disclaimer reminding students to always refer back to the model cards for the nuances in each model.}
  \Description{todo}
  \label{overview}
\end{figure}

We also offered a table summarizing the performances of the eight models to help comprehension and discussion (Figure~\ref{overview}). 

\subsection{The Persona Cards}
\begin{figure}[h]
  \centering
  \includegraphics[width=\linewidth]{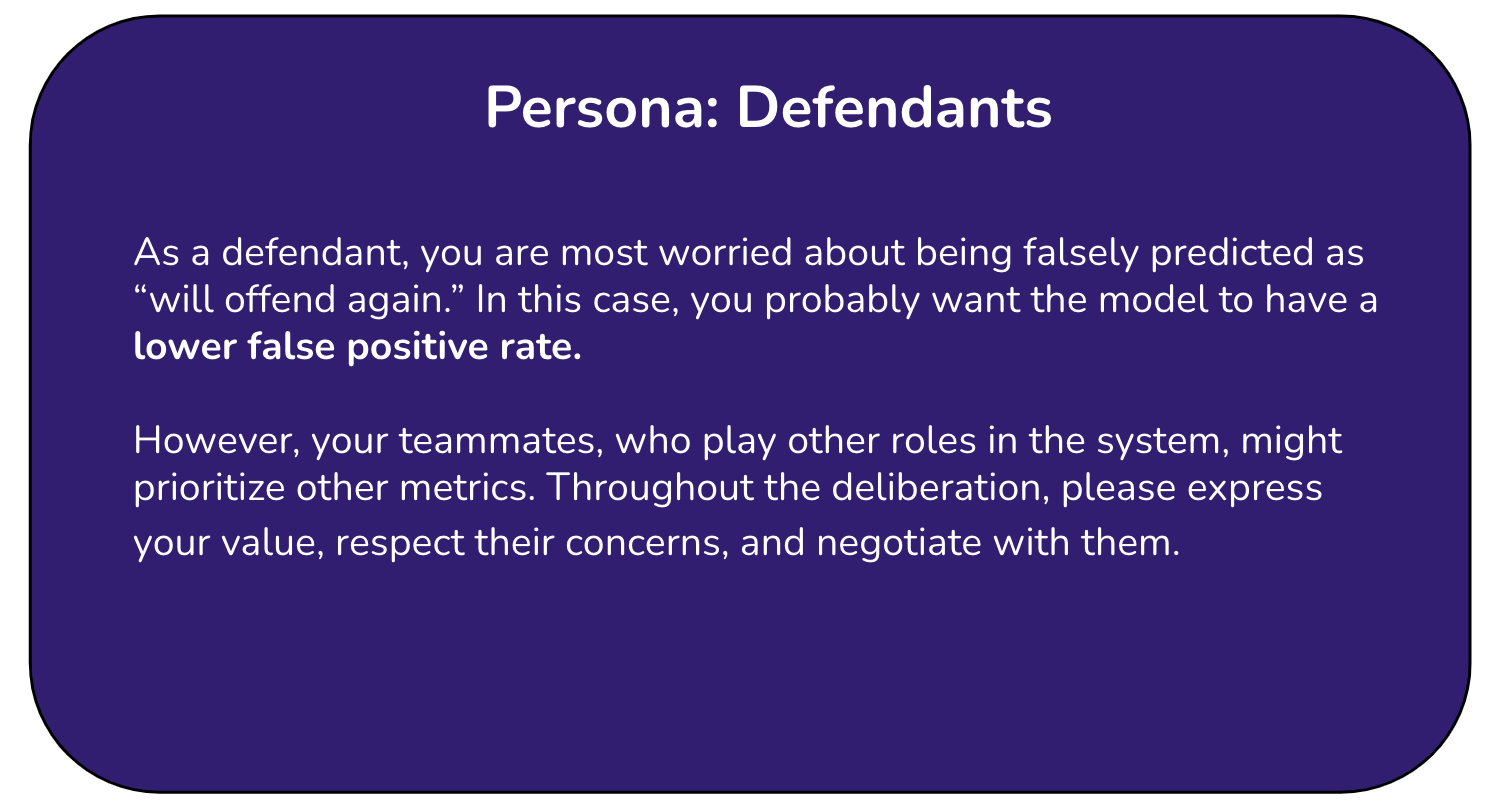}
  \caption{A Persona Card (Defendants).}
  \label{tab:Persona}
  \Description{todo}
\end{figure}

The second set of artifacts in the Value Cards toolkit is the Persona Cards. The Persona Cards are a series of cards that characterize different potential stakeholders and their prioritized values and interests in an algorithmic decision making system, with each corresponding to a number of specific Model Cards. They describe the backgrounds of each potential stakeholder involved in the system, the stakeholders' primary considerations when facing the outcome of the system, and some brief guidelines on how to engage in a productive discussion with other stakeholders (see Figure~\ref{tab:Persona}). 

The purpose of the Persona Cards is twofold. First, it offers a compact value overview of a high-stakes algorithmic system by showcasing the key stakeholders and their various perspectives. Second, each persona card offers the readers an access to a specific stakeholder's thinking process while the stakeholder is interacting with an ML-based decision-making algorithm. We hoped that students would take the persona card as a gateway to develop empathy towards the stakeholders and be ready to engage in the deliberation. 

In the case of recidivism prediction algorithm, we followed Narayanan's stakeholder mapping \cite{narayanan2018translation}, dividing potential social groups involved in the recidivism prediction case into four different groups of stakeholders: Judges, Defendants, Community Members, and Fairness Advocates, with each prioritizing one performance metric. Judges might want to prioritize increasing Accuracy when considering the design of a recidivism prediction system. Defendants might want to prioritize decreasing False Positive Rate as they are worried being falsely predicted as ``will offend again''. Community Members might want to prioritize decreasing False Negative Rate as they are mostly concerned about the safety of the community. Fairness Advocates might want to prioritize decreasing disparity as they want to minimize the differential treatment African American and White defendants. 

\subsection{The Checklist Cards} 
\begin{figure}[h]
  \centering
  \includegraphics[width=\linewidth]{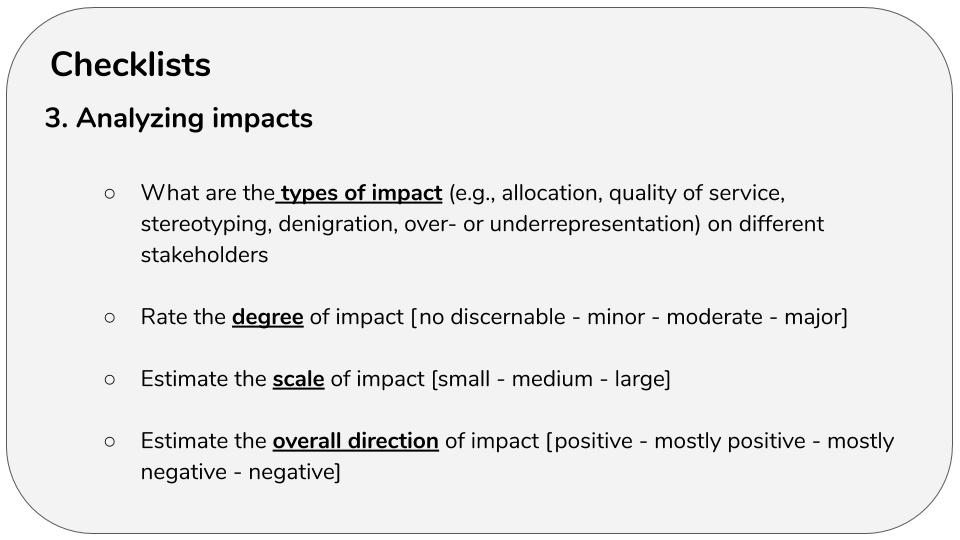}
  \caption{A Checklist Card (Analyzing Impacts).}
  \vspace{-10pt}
  \label{checklist}
  \Description{todo}
\end{figure}

The third set of artifacts in the Value Cards toolkit is the Checklist Cards (see Figure~\ref{checklist}). 

Inspired by Madaio et al.'s work on co-designing AI fairness checklist with industrial practitioners \cite{madaio2020co}, we consider the use of the Checklist Cards as a way to generate productive discussion, formalize ad-hoc processes, and empower individual advocates. 

We derived our Checklist Cards from previous work \cite{madaio2020co,UK2019,JK2019}, which were all designed as practitioner-facing toolkits to guide the ethical development and deployment of AI systems. Using these three checklists as our initial dataset, we performed affinity diagramming \cite{hanington2012universal} iteratively in our research group to cluster similar ideas, identify common themes, and combine different options.

Our final version of the Checklist Cards includes three subsets. The first is ``Understanding Societal Values in AI,'' which offers high-level points in considering social impacts of AI systems. The second is ``Identifying Stakeholders,'' which provides a starting point for students to think about who is at risk of experiencing impacts. The third set is ``Analyzing Impacts,'' which asks students to identify the type, degree, scale and overall direction of the impacts. 

\subsection{Limited Scope of the Case Study} 
We remark that the goal of our study is not to capture all unfairness issues in recidivism prediction, but to study the initial effectiveness of the Value Cards as an educational toolkit. Due to several limitations in carrying out the activity in a classroom meeting, we choose to focus on a restrictive set of model performance metrics, demographic groups, personas, and their associated priorities. In particular, throughout the development of the Persona Cards, we are aware that, in the real world, stakeholders have a myriad of identities -- race, gender, class, sexuality -- that shape their values, political interests, and interactions with a given algorithmic system \cite{10.1145/3134744}. However, we consider specifically scaffolding the stakeholder persona necessary in this early use of our approach in a classroom setting, as our student body is relatively homogeneous and many might not have sufficient domain knowledge. In our instruction, we remind students that each stakeholder's value offered in the Persona Cards should only be served as a starting point; and that students are welcome to and encouraged to enrich the background and celebrate stakeholders' intersectionalities. After the class activity, we also conducted an in-class reflection, where students reflected on issues not captured in the deliberation process.

\section{A Case Study Using the Value Cards Toolkit in Practice}
In this section, we describe a case study of using the Value Cards Toolkit in an authentic college classroom to facilitate student learning and deliberation about the social influences of machine learning algorithms. We designed a study in which we ask students to use the Value Cards Toolkit during a class meeting. In order to observe the effects of the Persona Cards and the Checklist Cards respectively, we designed four conditions of using different components of the toolkit. We also employ a mixed-methods approach, using both quantitative and qualitative measures to assess the learning outcomes for students. 

\subsection{Study Context}
We conducted our study in the Human-AI Interaction class at Carnegie Mellon University in Fall 2020. \footnote{Course website is available at https://haiicmu.github.io/} The goal of the class is to introduce students to ways of thinking about how Artificial Intelligence will and has impacted humans, and teach students approaches to design AI systems that are usable and beneficial to humans and the society. The intended learning goals of the Value Cards Toolkit align very well with the learning objectives of the course, making the course a natural case study candidate for our Value Cards toolkit. We conducted the study in September 2020, at the beginning of the Fall semester, when students were just starting to learn about machine learning concepts and performance metrics. We have acquired IRB permission from our institution under number STUDY2020\_00000356 to perform the study and also obtained students' consent on using their anonymized data.

\subsection{Study Design and Implementation}

During COVID-19, class meetings were held remotely via Zoom. Students were asked to complete a 5-step activity using the Value Cards Toolkit during one class meeting. All students were assigned to teams and each team was assigned to one of the four conditions. The detailed instructions and the Values Cards toolkit were given to the students in digital version through Canvas \cite{Canvas}. 
The study has the following five steps: (1) Students complete an individual quiz. (2) Students read Model Cards and make individual model selections. (3) Students are assigned to teams of 4 and have team discussions via Zoom breakout rooms around. The 4 students in a team collaboratively write a short proposal. (4) Students complete a post-survey individually, and the instructors hold a subsequent reflection session to collect feedback. (5) Students complete a post-quiz 7 days after the initial in-class activity. Students were given the Model Cards at Step 2, the Persona Cards at Step 2 if applicable, and the Checklist Cards at Step 3 if applicable. See Table~\ref{tab:Conditions} for details on conditions and random assignments.

\begin{figure}[h]
  \centering
  \includegraphics[width=\linewidth]{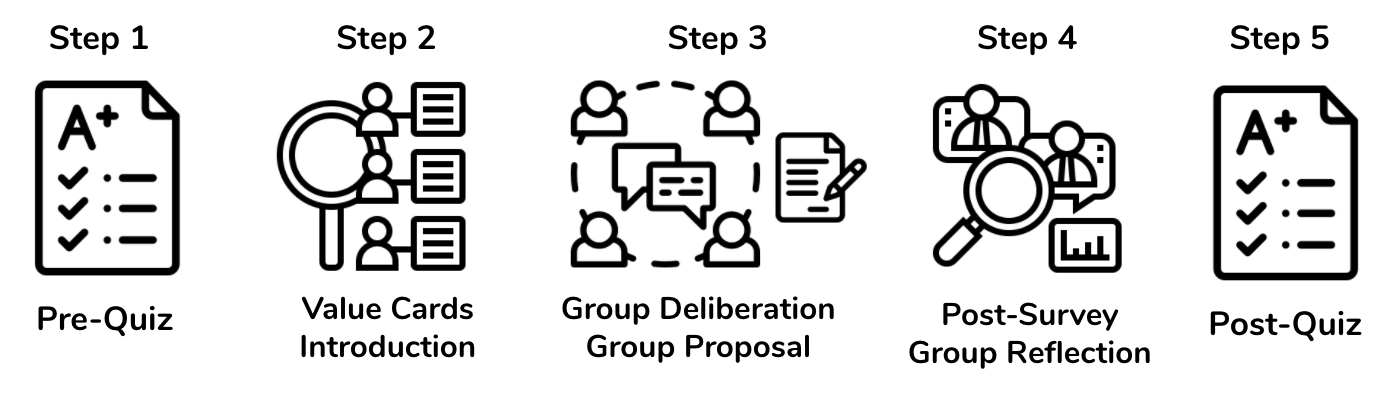}
  \caption{Five Steps of Study Design}
  \Description{todo}
\end{figure}

d
\begin{table}[b!]
\centering
\renewcommand{\arraystretch}{1.3}
\begin{tabular}{p{2cm}|p{3.5cm}|p{2cm}}

\textbf{Condition Names}& \textbf{Condition Details}& \textbf{Room Assignments}                       \\\specialrule{.1em}{.05em}{.05em}
\textbf{Baseline}  & Model Cards. \textbf{\textit{No}} Persona Cards. \textbf{\textit{No}} Checklist Cards & 1, 2, 3      \\\hline

\textbf{Checklist Only} & Model Cards and Checklists Card. \textbf{\textit{No}} Persona Cards. & 4, 5, 6      \\\hline

\textbf{Persona Only} & Model Cards and Persona Card. \textbf{\textit{No}} Checklist Cards & 7, 8, 9, 10      \\\hline

\textbf{Checklist and Persona} & Model Cards, Persona Cards, and Checklist Cards & 11, 12, 13, 14            
\end{tabular}

\caption{This table displays our four study conditions' names, details, and the corresponding room assignments.}
\label{tab:Conditions}
\end{table}

\subsection{Outcome Evaluations}
We designed and administered four outcome evaluations to assess student learning. This includes a matched pre- and post-quiz, a post-survey, and an open-ended written group proposal. In addition, we anticipate that student learning will be observed in the deliberation process facilitated by the toolkit. We further analyze the student teams' deliberation process to evaluate student learning along with all three learning objectives. In this section, we introduce how these four outcome evaluations are designed and implemented. The alignment between the learning objectives and the outcome evaluations is shown in Table ~\ref{tab:mapping-objectives-outcomes}.

\begin{table}[b!]
\centering
\renewcommand{\arraystretch}{1.3}
\begin{tabular}{p{5cm}|p{2.8cm}}
\textbf{Learning Objectives}& \textbf{Outcome Evaluations}                       \\ \specialrule{.1em}{.05em}{.05em}
1. Understand the technical definitions and trade-offs of performance metrics in machine learning, and apply them in real-world contexts. & Pre- and post- quizzes; Deliberation       \\\hline
2. Understand the importance of considering diverse stakeholders' perspectives in the development and deployment of machine learning systems.                                &  Post-survey and Reflection; Group proposal; Deliberation 
\\\hline 
3. Being able to communicate, negotiate, and synthesize the perspectives of diverse stakeholders when algorithmic decisions have consequential impacts.  & Group proposal; Deliberation           
\end{tabular}

\caption{This table displays the learning objectives of the Value Cards toolkit and the corresponding outcome evaluations we use in the case study to assess student learning.}
\label{tab:mapping-objectives-outcomes}
\end{table}

\subsubsection{Quiz on Conceptual Understanding}
We administered a pre-quiz and a delayed post-quiz to assess Learning Objective \#1. We gave students a toy dataset and a prediction task and ask students to apply the performance metrics on this dataset. We used the criminal recidivism prediction task in the pre-quiz and the loan application prediction task in the post-quiz. The quiz contained questions about the basic understanding and computation of performance metrics and questions about evaluating the trade-offs between different performance metrics.
At the end of the pre-quiz, we also inserted a subjective question asking about students' considerations of performance metrics when designing machine learning algorithms. 
 The pre- and post-quizzes were designed by two experienced instructors of the topic. We pilot tested the quizzes with two students iteratively to improve clarity. Student performance data were directly collected via Canvas.


\subsubsection{Post-survey and In-Class Reflection}
We designed a post-survey to evaluate Learning Objective \#2. In the survey, we first included the last subjective question in the pre-quiz again. This allows us to compare whether students' considerations of metrics changed after using the Value Cards toolkit. We then included 4 likert-scale questions asking students to what extent they understand other stakeholders' perspectives in recividism prediction. We also inserted an open-ended question for students to elaborate on their responses. 

We included a series of self-evaluation questions and a section on demographics. 
Following the individual post-survey, the instructors hosted a reflection session to collect students' takeaways and feedback of the activity via a Zoom class meeting. 
To make sure students offer authentic feedback and to protect student privacy, students' responses from the post-survey and the in-class reflection were collected and recorded anonymously.

\subsubsection{Group Proposal}
In order to assess Learning Objective \#3, we asked each team to collaboratively write a group proposal following their deliberation facilitated by the Value Cards toolkit. We gave students the following instructions: ``Now your job is to discuss with your teammates on which algorithm you think is producing the best outcomes and would recommend for the policymakers to use. 
If you’re able to agree on a model, please share your reasons. If you’re not able to reach a consensus, please also share your reasons.'' Student teams wrote the proposals in Google docs, and we analyzed the quality of the proposals to evaluate student learning.

\subsubsection{Process Evaluation of Group Deliberation}
In addition to the three outcome evaluations we explicitly designed and administered, we also anticipated that student learning could be observed from their group deliberation facilitated by the Value Cards toolkit. The deliberation process was audio-taped and transcribed. We analyzed the group deliberation to evaluate whether there is evidence of student learning along with all three learning objectives.

\subsection{Participant Demographics}
In total, 62 students participated in the study, we removed the data from 6 students who chose to opt-out of the data collection. Our final dataset contained 56 responses. It included 55\% female participants, 43\% male participants, and 2\% prefer not to say.  46\% of the sample were undergraduate students while 54\% of the sample are pursuing a graduate degree. 80.3\% of the participants identified as Asian, 3.6\% as Black or African American, 12.5\% as White and 3.6\% as Hispanic, Latino/a/x, or Spanish Origin. The participants ranged in age from 18-44 (64 \% 18-24, 32 \% 25-34, 4 \% 35-44). 

Among the 56 students who participated in the study, two students missed the pre-quiz, and one student missed the post-quiz, leaving us 53 data points to conduct the pre-quiz/post-quiz analysis. All 14 rooms uploaded their discussion audios and group proposals. All students submitted the individual model selection; 55 students submitted their post-survey. 

\subsection{Data Analysis}
We adopted a mixed-methods approach for data analysis. First, we performed a quantitative analysis on students' learning gains from pre- to post- quizzes, their change on metrics consideration after the activity, and their understanding of diverse perspectives as shown by the multiple-choice questions in the post-survey. Second, we performed a qualitative analysis on the group deliberation, open-ended responses in post-survey and in-class reflection, and the group proposals to give a richer and more comprehensive view of student learning and their demonstration of understanding.

\subsubsection{Quantitative Analysis}
The main outcome variables are the quiz scores and metrics selections (i.e., their responses to the question of \emph{``For any algorithmic decision-making system, what metrics below do you think are the most important ones to consider in tuning the algorithm?''}). Note that for each participant, we collected their quiz scores and the numbers of the metrics they chose twice, before and after the Value Card activity. This allows us to evaluate whether the Value Card activity helped improve students' understanding of the concepts, and reduce the reliance on single performance metrics in the development of machine learning algorithms (learning objective 1 and 2). We conducted a paired t test to examine whether the pre-post change were statistically significant. To examine the difference across conditions, we ran regression analysis with the post-activity quiz scores or metrics choices as dependent variable, conditions as independent variables, and pre-activity quiz scores or metrics choices as control variables.  

\subsubsection{Qualitative Analysis}
The qualitative datasets used in this study include group proposals, open-ended questions in the post-survey, in-class reflection and group discussions. The in-class reflection and group discussions were audio-recorded and transcribed using otter.ai \cite{otter}. 
In this case study, our goal is to examine student learning along with the three objectives as shown in Table~\ref{tab:mapping-objectives-outcomes}. 
We started from inductive coding \cite{braun2006using} to extract codes that show evidence of learning and developed a codebook. In total, we summarized 6 main codes and 38 subcodes. In the end we used a deductive coding approach, applying the codebook on the entire dataset to present evidence of learning along with the three objectives. Our approach builds upon and differs from existing grounded theory \cite{charmaz2014constructing} and thematic analysis \cite{braun2006using} approaches in that the goal of the inductive coding is not to find out emerging themes and summarizing themes, rather to develop a codebook that serves as a vehicle for the subsequent deductive coding to undercover evidence and insights of student learning along with the three objectives. 



\section{Findings}
In general, our class activity suggests that the use of the Value Cards toolkit is promising in achieving the three learning objectives outlined in Table 2. However, our results also suggest that there are a number of pros and cons that need to be considered when using the different variants of the Value Cards approach.

Below, we report our findings through the quantitative analysis comparing the two quizzes, and the qualitative analysis aggregating the deliberation transcripts, group proposal, in-class reflection and open-ended questions in the quizzes and post-survey. We also present students' critiques towards the different variants of the toolkit, especially the Persona Cards and the Checklist Cards.

We use ``R'' to represent discussion room; and ``P'' to identify specific participants. We don't include IDs for anonymous surveys and in-class reflection. 

\subsection{Achieving Learning Objectives}
\subsubsection{Students improved their understandings of the technical definitions and trade-offs of performance metrics in machine learning, and were able to apply them in real-world contexts (Objective 1)}\

\begin{figure}[h]
  \centering
  \includegraphics[width=\linewidth]{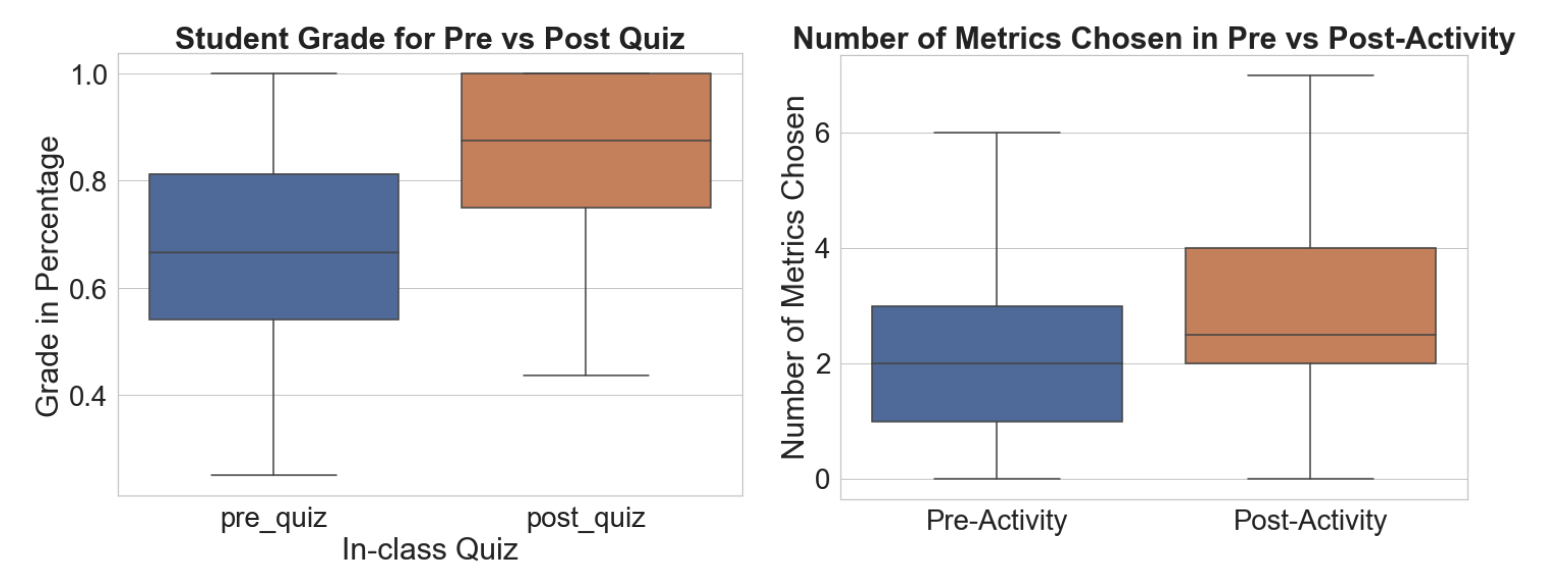}
  \caption{Pre- and Post-Activity Quiz Scores and Numbers of Metric Chosen}
  \label{quan}
  \Description{todo}
\end{figure}

We used pre- and post-quizzes to measure the learning effects on conceptual knowledge. As a reminder to our readers, the pre-quiz (Step 1) happened at the very beginning of our in-class activity, after the instructors gave an introduction about the technical definitions in a previous lecture. The post quiz took place 7 days after the activity. The two quizzes are both designed to assess students' understandings of the technical definitions and trade-offs of performance metrics in machine learning.

In general, students performed better in the post-quiz (avg = 0.84, std = 0.14) than in the first quiz (avg = 0.68, std = 0.18). The difference is significant ($p<0.01$ in the paired t-test) (see Figure 7). This suggests that the use of the Value Cards toolkit significantly improved students' comprehension of the technical definition and trade-offs of performance metrics. 


In a traditional machine learning course, the definitions of False Negative Rate and False Positive Rate are usually presented by the instructors in a mathematical formalization without engaging students within a specific social context.   
In our study, the immersive deliberation procedure served as a friendly yet effective environment for students to proactively apply the technical terminologies they've learned into a real-world case study. Students helped each other consolidate their understanding of technical ML terms and the metrics' trade-offs. 

For example, one student in Room 5 (Checklist Only) explained the False Positive Rate in the context of recidivism prediction as: 
\begin{quote}
    \emph{It's having people who are actually not going to commit another crime test positive. It's like just calling them out as guilty even if they have no plan to commit any crime.} (R5, P17)
\end{quote}

By observing the various metric performances of the eight models we offered, students deepened their understanding of the trade-off between different metrics. In Room 13 (Checklist and Persona), students had a conversation as below:
\begin{quote}
    \emph{I mean, I'm taking a higher false positive rate, but it's definitely a big step to lower the false negative rate...} (R13, P51)
\end{quote}

Towards the end of the deliberation in Room 2 (Baseline), P6 commented that
    \emph{``I wish we did this part of the exercise before the quiz (Step 1) because now I understand everything way better.''}

\subsubsection{Students understood the importance of considering diverse stakeholders' perspectives in the development and deployment of machine learning systems (Objective 2).}\

In both the pre-quiz and the post-survey, we asked the students questions regarding the performance metrics they want to prioritize in designing a recidivism prediction algorithm. Our quantitative analysis showed that students selected more metrics after the activities (avg = 2.8) compared to before the activities (avg = 2.3); the difference is significant ($p = 0.04$), which suggested that our activity helped the students better recognize there are diverse social values embedded in the algorithmic system (see Figure 7).

In our post-survey, we asked students in all conditions if they understood the perspectives of the judges, defendants, community members, and fairness advocates after the activity. The average score is 5.6 out of 7. We report the variances in perspective taking across different conditions in section 5.2.2.

We noticed a large population of our students endorsing the model with the highest overall accuracy at the beginning of the deliberation. However, after the group discussion, they started to recognize that overall accuracy fails to capture the nuance of social consequences the machine learning systems might have. For example, one student commented in the post-survey:
\begin{quote}
 \emph{(The activity) made me reconsider my previous perspectives that models can be "ranked" on an absolute basis by their accuracy. Although accuracy is a great metric, it is extremely general and fails to capture the nuances of societal costs from using a model with high accuracy at all costs.}
\end{quote}

One student commented why they selected more performance metrics in the post-quiz,
\begin{quote}
 \emph{After hearing everyone's arguments for their point of view, I kind of began to think all of the aspects are very important in tuning an algorithm. The costs are very high no matter what, so it is important to view them all as equally prioritized.}
\end{quote}

Another student in Room 9 (Persona Only) also commented referring to the current events happening in the real world,
\begin{quote}
 \emph{This is why disparity is very important. Because if there are big differences between the prediction results in White Americans and African Americans, it will hurt people, especially recently we have seen much news about discrimination. So I think we cannot hurt people anymore.} (R9, P36)
 \end{quote}








In sum, we observed an general pattern emerging in our data across post-survey, discussion, and group proposal that students started to recognize the importance of considering diverse social values in building machine learning systems. 

\subsubsection{Students were able to communicate, negotiate, and synthesize the perspectives of diverse stakeholders (Objective 3).}\

We further observed that students started to consider, negotiate and reflect on the diverse and competing stakeholders' perspectives in their group discussion. For example, students were able to assertively advocate for a stakeholder's interest,

\begin{quote}
    \emph{I'm concerned about the safety of my neighborhood. The false-negative rates for those cards you mentioned are really high. It seems like we will disagree with each other.} (R8, P29)
\end{quote}

When facing conflicts, students actively negotiated and compromised in order to reach a common ground.  

\begin{quote}
 \emph{The models are extreme in one way or another. It either has super high accuracy or very low disparity. To satisfy everyone, we need to reach some common ground.} (R14, P55)
 \end{quote}
 

During the negotiation, students in general showed respect to other's perspectives and values. For instance, P50 in Room 13 (Checklist and Persona) made the following comment after another student expressed concerns about a model's high false positive rate:

 
 
 
 
\begin{quote}
 \emph{Yeah we should all consider other people's perspectives... Then after listening to what he has talked about, I think it's bad for us to have like a 50\% of false positive rate.} (R13, P50)
 \end{quote}

For group proposals, 10 out of 14 teams reached a consensus and picked up a model. Students within these 10 teams all supported their group model choices with well-thought-out arguments. For example, students in Room 9 (Persona Only) started their proposal with careful consideration and balancing of all perspectives: 
\begin{quote}
 \emph{We started with the principle that within the United States you are innocent until proven guilty. As such, though we each had varying perspectives, we felt that the false positive rate should be an attribute strongly considered.}
\end{quote} 

There were 4 out of 14 teams that did not reach a consensus. Even though they failed to achieve consensus, their group proposals reflected their thoughtfulness in the decision making. For example, in Room 6 (Checklist Only), students didn't reach consensus because of the inevitable trade-offs between FPR and FNR.








\subsection{Different Combinations of the Toolkits}

\subsubsection{Pros and Cons of Using the Persona Cards.}\
We observed some potentially promising effects of using the Persona Cards in achieving our learning objectives. For conceptual knowledge, our quantitative analysis suggested that students in the persona conditions performed marginally better than the students in the non-persona conditions ($p = 0.07$) in the second quiz, controlling their performance in the first quiz. Having persona cards does not significantly impact students' choices of performance metrics. 



We see clear evidence from the qualitative data that having Persona Cards help some students quickly gain empathy towards the stakeholders, and articulate their values.
 
\begin{quote}
\emph{As a defendant, I'm really worried about being falsely predicted as high risk ... like a repeat offender ... I don't want my life being destroyed just for that reason. 
(R13, P52)}
\end{quote}

In addition to the perspectives we offered in the Persona Cards, students also enriched the personas by leveraging their own understandings about the stakeholders and the criminal system:

 
 \begin{quote}
 \emph{Besides my persona, I think we also need to look at the societal costs... I think it's really dependent on the types of crimes that you're diagnosing. For example, if you're diagnosing crimes like petty theft, or shoplifting, things like that, then a false negative is not a very high cost for society to pay in exchange for overall accuracy.} (R12, P46) 
 \end{quote}

In general, Persona Cards enabled students to simulate different stakeholders, which could potentially scaffold more effective deliberations. In post-survey,  one student commented,
\begin{quote}
\emph{I think that discussing the perspective of defendants while trying to choose a model really gave me a moment to step into the shoes of the defendants and understand what the impacts of different decisions could have on them and others.}
 \end{quote}

In comparison, teams in non-Persona conditions sometimes struggled with identifying stakeholders in discussion. One student in the "Checklist Only" condition commented in the post-survey: 
\begin{quote}
 \emph{I think in general, we are not given the instructions to think about the aspect of different stakeholders... but if there is a little bit more hint ... right now I was only thinking about race bias and defendants"}
 \end{quote}

However, the use of the Persona Cards also has its limitations. For example, during the post-activity reflection session, when asked whether or not the Persona Cards helped their discussion, some students commented that they \emph{``can't really put myself in my persona's position''}. Some think the design of the Persona Card is too simplified and is \emph{``missing some value''}.  

Interestingly, several students also mentioned that having a Persona Card made them more stubborn. One student said,

\begin{quote}
 \emph{Persona Card made me defensive. I'm fighting for my persona. I'm not considering other personas' perspectives as much as I would like to.}
 \end{quote}

We did observe several overly intense debate among two personas in some of the discussion transcripts. 

Some reasonable approaches to tackle these issues could be presenting the personas narratively in a story instead of in separate cards, or having the students collectively create their own set of Persona Cards. We suggest additional potential alternative designs of the Persona Cards in the Discussion session.

\subsubsection{Pros and Cons of Using the Checklist Cards.}\
The use of the Checklist Cards also has both positive and negative effects. 

On the one hand, we noticed positive signs of using the Checklist Cards in scaffolding the discussion process, in particular, to generate productive discussion around the diverse social values in the development and deployment of machine learning systems. For example, Room 4 (Checklist Only) strictly followed the framework we offered in the Checklists Card during the deliberation. Students in Room 4 started with identifying societal values in the context of recidivism prediction,
then identified a set of relevant stakeholders, and finally analyzed the impacts of different models' outcomes. In their Group Proposals, they started by mapping out various stakeholders directly or indirectly impacted by the recidivism prediction system, including ``releasable criminals,'' ``inmates,'' ``guards,'' ``taxpayers,'' ``politicians,'' etc. and listed related harms and benefits caused by the system correspondingly.


Our results also suggest that the use of the Checklist Cards alone, without the Persona Cards, might not help students understand the different stakeholders’ perspectives. In our post-survey, we asked whether they understood the perspectives of the judges, community members, defendants, and fairness advocates after the activity. Students in the ``Checklist Only'' condition responded lowest (5.0 out of 7) compared to the other three conditions (5.8 for the ``Baseline'' condition, 5.7 for the ``Persona Only'' condition, 6.3 for the ``Checklist and Persona'' condition). The difference is significant (\emph{p}=0.04).

One possible explanation is that for the Checklist Only condition, students spent more time discussing the checklist, and then had less time discussing the perspectives of potential stakeholders. This aligns with the over-scripting \cite{dillenbourg2002over} literature, which suggests that scripts for collaboration might be ineffective when they disrupt the team's original conversation flows.

Interestingly, when the checklists and personas were presented to the students simultaneously (``Checklist and Persona'' condition), students' discussions were driven by the personas more than the checklist. Specifically, discussions in Room 11, 13, and 14 did not refer to the Checklists in their discussion at all. Room 12 had one student who mentioned the checklist only once, to confirm that they were in the right track:

\begin{quotation}

    \emph{I think this is a good sign for us to go into part two (on the checklist), which is like, who actually matters, like who the stakeholders are?} (R12, P46)

 \end{quotation}

During the post-activity reflection session, students mentioned that they didn't check out all the points on the Checklists card due to time constraint. One student also complained that the checklists were too verbose, and the team members were too busy arguing for their personas, so they ignored the checklist during the deliberation. One potential way to mitigate this drawback could be revealing the Checklist Cards in an earlier stage of the activity, so that students have abundant time to read through and internalize the checklists. 

\section{Discussion}
In this research, we proposed a deliberation-driven toolkit -- the Value Cards -- to inform computer science students about the social impacts of machine learning-based decision making systems. We conducted a classroom activity to evaluate the effectiveness of our approach. The ``proof-of-concept'' case study demonstrated that the toolkit is promising in helping students understand both the technical definitions and trade-offs of performance metrics and apply them in real-world contexts. 

Our findings also suggested nuances in using different combinations of the Value Card set. Our results suggest that, although Persona Cards helped some students quickly gain empathy towards the stakeholders, some other students had difficulty in putting themselves in the persona's positions. Persona Cards could also make some students more stubborn and less accepting of other personas' perspectives. Checklists can help scaffold the discussion; however, spending too much time discussing the checklist itself might disrupt the team's original conversation flow. When the checklists and personas were presented to the students simultaneously, students’ discussions were likely to be driven by the personas more than the checklist. In sum, future users might want to tailor the use of the Value Cards toolkit for their specific settings or domains.

Interestingly, during the study, we also saw students start to think about things in more ways than what was specifically offered in the toolkit. For example, students started to identity more stakeholders in addition to the ones offered in the Persona Cards, in R4, P15 commented, \textit{``I'm thinking about tax payers, they pay the money to support the system. They're like also indirect stakeholders.''} 

They also started to think about better ways of integrating humans in the loop. For example, in R2, P7 mentioned, \textit{``in my head, there will be another layer, you know, you have another person checking this process and the model's job is to make that person feel easier.''}




As an initial effort, there are a number of limitations of our ``proof-of-concept'' study that are important to mention.

First, in this work, we utilized a top-down approach to create the Persona Cards, specifically designating four different stakeholders and their competing values, based on the stakeholder mapping offered in \cite{narayanan2018translation}. We are aware that this practice might create abstractions and even ``objectified assumputions'' about the diverse social values involved in the recidivism prediction system \cite{costanza2020design}. We consider this top-down approach as a necessary starting point to generate productive discussion within a rather homogeneous student body. Future work is needed to explore bottom-up approaches in creating the Persona Cards via real-world community engagement. For example, we can borrow the ``create your own'' cards practice from the design of the Envisioning Cards \cite{friedman2012envisioning} and ask our participants to design and generate their own Persona Cards. 

Second, it is also important to note that deliberation is just one mode of collective decision making. Past scholarship has offered important critiques to both the concept of deliberation and its practices (e.g., \cite{fraser_rethinking_1992}). In this work, we used the deliberation process primarily as a way to foreground the \textit{social} aspects of seemingly neutral technical process of adjusting system criteria in algorithm design. Future work is needed to explore alternative methods. For example, Mouffe's theories of agnostic pluralism \cite{mouffe1999deliberative} remind us of the importance of contentious expression as an alternative and complement to rational deliberation. 

Third, in this work, we evaluated the initial effectiveness of our approach via a classroom activity. Future work is needed to explore the utility of the Value Cards in other educational settings, for example, in large cross institutional study, or to complement the existing technical training for ML practitioners in both the public and private sectors. 
In addition, we can also explore the use of the Value Cards toolkit in real-world community engagement around algorithmic decision making systems (e.g., in community workshops), with lay people-oriented design of the Model Cards and Checklist Cards.

Fourth, similarly, in this work, our case study explored the design and use of the Value Cards in the context of recidivism prediction algorithm. Future research is needed to replicate and validate our methods and findings in other decision-making contexts. 

Finally, due to the constraints posted by the ongoing Covid-19 pandemic, our class activity was held entirely on Zoom, which means the Value Cards were all presented to study participants in the digital formats. This suggests opportunities for future research to test the use of the physical cards in real-world interactions. 

\section{Conclusion}
In this paper, we present the Value Cards, a deliberation-driven toolkit for informing computer science students and practitioners the social impacts of different performance metrics and their trade-offs. We documented an early use of the Value Cards in a college-level computer science course and reported the initial effectiveness of our approach. Our results suggested that our approach is promising in helping the students comprehend, negotiate and reflect on the social values embedded in machine learning-based algorithmic systems. 

\section{Acknowledgments}
We thank Pranav Khadpe and our other anonymous reviewers and colleagues at the HCII at Carnegie Mellon University for their feedback. This work was supported by the National Science Foundation (NSF) under Award No. IIS-2001851, CNS-1952085, IIS-2000782, and DLR-1740775, the NSF Program on Fairness in AI in collaboration with Amazon under Award No. IIS-1939606, an Amazon Research Award, and a J.P. Morgan Faculty Award.



\bibliographystyle{ACM-Reference-Format}
\bibliography{FAcct2021}


\end{document}